\documentclass[a4paper,11pt]{article}
\usepackage{pos}

\newcommand{\SO}{\text{SO}}
\newcommand{\SU}{\text{SU}}
\newcommand{\U}{\text{U}}
\newcommand{\Sp}{\text{Sp}}

\def\gsim{\raise0.3ex\hbox{$\;>$\kern-0.75em\raise-1.1ex\hbox{$\sim\;$}}}
\def\lsim{\raise0.3ex\hbox{$\;<$\kern-0.75em\raise-1.1ex\hbox{$\sim\;$}}}
\def\ptmiss{p_{T}^{\rm miss}}

\title{LHC phenomenology of unusual top partners
in composite Higgs models}
\author*{Werner Porod}

\affiliation{Institut f\"ur Theoretische Physik und Astrophysik, Uni W\"urzburg,
Emil-Hilb-Weg 22,
D-97074 W\"urzburg, Germany}

\emailAdd{porod@physik.uni-wuerzburg.de}

\abstract{Composite Higgs models with a fermionic UV completion can contain additional colored
states beside the usual top-partners. We focus here on a model which contains in addition 
SU(3) color octet top partners as well as color singlet ones. The latter can in principle serve
as a dark matter candidate. 
We consider a particular composite Higgs model which contains SU(3) color octet top partners besides the
usually considered triplet representations. Moreover, color singlet top partners are present as well which can in principle serve as dark matter candidates. We investigate the LHC phenomenology of these unusual top partners. Some of these states could  at first glance be confused with gluinos predicted in supersymmetric models.
}

\FullConference{%
  Corfu Summer Institute 2021 "School and Workshops on Elementary Particle Physics and Gravity"\\
  29 August - 9 October 2021\\
  Corfu, Greece
}

\begin{document}
\maketitle

\section{Introduction}

The Standard Model (SM) is quite some success story which while been challenged in several
experiments shows are remarkable agreement between theory and experiment.   
The discovery of a Higgs-like scalar resonance at the LHC \cite{ATLAS:2012yve,CMS:2012qbp} has
has completed this framework particle-wise. However, the lightness of the observed boson,
whose properties agree with the SM prediction, has
 materialized long-standing questions on the SM as the ultimate theory of particle interactions. 
Why is the Higgs boson mass insensitive to the scale of new physics (Planck mass)? 
Is there a dynamical origin for the spontaneous breaking of the electroweak (EW) symmetry?
Is it indeed an elementary scalar particle?

These questions can be addressed by tying the properties of the Higgs-like scalar to those of
fundamental fermionic states, which do not suffer from quantum sensitivity to large scales.
The two time-honoured avenues realizing this idea are supersymmetry (SUSY) and compositeness. In SUSY
 scalars are associated with fermions via a new symmetry extending Poincar\'e invariance of 
particle interactions. In compositeness models scalars emerge as resonances of underlying bound fermions.
In this contribution we follow the composite avenue. The basic idea is inspired by QCD
but realized with an extended symmetry of the condensing theory to allow for a misaligned vacuum \cite{Kaplan:1983fs}. The large top Yukawa coupling can be explained via the mechanism
of partial compositeness \cite{Kaplan:1991dc}. 
At the price of a moderate tuning, this class of models features a limit where a light Higgs-like state 
emerges as a pseudo-Nambu-Goldstone boson (pNGB). This idea got boosted in the early 2000s thanks to 
the holographic principle \cite{Contino:2003ve} linking near-conformal theories in 4 dimensions to 
gauge/gravity theories in 5 dimensions on a warped background. 
Based on the symmetry breaking pattern $\SO(5)/\SO(4)$ a `minimal' model was proposed 
\cite{Agashe:2004rs}, where the number of pNGBs matches the 4 degrees of freedom of the 
SM Higgs field. The phenomenology of this model has been widely explored, see e.g.~the reviews \cite{Contino:2010rs,Bellazzini:2014yua,Panico:2015jxa}. 

The minimal model and variations thereof are a useful templates to understand the phenomenology 
of a composite pNGB Higgs. However, from the point of view of an underlying gauge-fermion theory
\`a la QCD the symmetry breaking pattern $\SO(5)/\SO(4)$ cannot be obtained as the global symmetry
group of the underlying fermions is unitary \cite{Cacciapaglia:2014uja,Cacciapaglia:2020kgq}.
The minimal breaking patterns for a composite Higgs which arises from models with underlying fermions 
are $\SU(4)/\Sp(4)$, $\SU(5)/\SO(5)$ or $\SU(4)\times \SU(4) / \SU(4)$ 
\cite{Dugan:1984hq,Galloway:2010bp,Cacciapaglia:2014uja}. These models 
predict additional pNGBs 
besides the usual Higgs doublet. Moreover, models featuring top partial compositeness 
\cite{Barnard:2013zea,Ferretti:2013kya,Vecchi:2015fma} require additional QCD-charged underlying 
fermions in order to allow for fermionic bound states with the same quantum numbers as the SM top quark. The underlying colored fermions condense like their electroweak counter parts yielding
additional colored pNGBs in the low energy effective theory \cite{Ferretti:2016upr}.

The presence of the additional pNGBs generates new decay channels for the top partners \cite{Bizot:2018tds} besides the ones usually considered in direct searches at the LHC. These can substantially shift the mass
bounds for these states for which we refer to \cite{Banerjee:2022xmu} for a summary. 
These models do not only have an extended pNGB sector but usually contain also additional baryonic
states which do not transform as $\SU(3)_c$ triplets. We will demonstrate some generic features 
focusing on one specific model, first proposed in ref.~\cite{Ferretti:2013kya} and dubbed M5 in 
ref.~\cite{Belyaev:2016ftv}. It belongs to a class of models, comprising 12 candidates, where the top partners 
arise as chimera baryons made of two different species of confining fermions. A peculiar feature of the model 
M5 compared to the others is that the baryon spectrum contains color-octet fermionic states. They are predicted to be among the lightest top partners an, thus, play the 
leading role in the LHC phenomenology of this model 
\cite{Cacciapaglia:2021uqh}. Moreover, the pNGB sector contains a color-triplet with 
the charge of a right-handed stop as well as a color octet. In addition color-neutral 
baryons are predicted. We will see that the spectrum and phenomenology of the model M5 shows surprising similarities to supersymmetric models. 
Finally, the properties of the confining gauge dynamics, based on $\Sp(4)$, is being studied on the Lattice 
with promising results~\cite{Bennett:2017kga,Bennett:2019jzz,Bennett:2019cxd}.
Complementary information on the mass spectrum and decay constants of the composite states can also be 
obtained using Nambu-Jona-Lasinio models \cite{Bizot:2016zyu} 
or holographic techniques \cite{Erdmenger:2020lvq,Erdmenger:2020flu}.

\section{Model aspects}

The M5 model has an $\Sp(4)$ hyper-color gauge group with 5 Weyl fermions $\psi_i$ in the antisymmetric and 6 
Weyl fermions $\chi_j$ in the fundamental representation as underlying particles of the composite sector. 
This fermion sector exhibits an $\SU(5)\times \SU(6)$ global symmetry. Its electroweak sector has 
been investigated in \cite{Agugliaro:2018vsu,Banerjee:2022izw,Banerjee:2022xmu}.
The chiral condensates $\langle \psi\psi \rangle$ and $\langle \chi\chi \rangle$ spontaneously 
break the global symmetry to the stability group $SO(5)\times Sp(6)$. 
The SM color group $\SU(3)_c$ is realized as a gauged $SU(3)$ subgroup of $\Sp(6)$, while the weak 
group $\SU(2)_L$ is a gauged subgroup of $\SO(5)$, which also contains a custodial subgroup 
$\SO(4) \sim \SU(2)_L \times \SU(2)_R$. Moreover, the $\U(1)_Y$ hypercharge $Y = T^3_R + X$ is a gauged linear 
combination of the diagonal generator of $\SU(2)_R\subset \SO(5)$ and $\U(1)_X\subset \Sp(6)$. In addition, 
the model contains two global abelian symmetries $\U(1)_\chi$ and $\U(1)_\psi$. One linear combination of 
these $\U(1)$ factors is $\Sp(4)$ anomaly free, and the spontaneous breaking by the condensates yields a pNGB, 
while the would-be pNGB associated to the orthogonal $\U(1)$ combination is expected to receive a mass 
through the $\Sp(4)$ anomaly.

This model contains  three classes of pNGBs:
\begin{enumerate}
 \item A SM singlet pNGB $a$ from the $\Sp(4)$ anomaly-free spontaneously broken $\U(1)$ which is expected to
  be light. It couples axion-like to SM particles and we refer to \cite{Belyaev:2016ftv,Cacciapaglia:2019bqz,BuarqueFranzosi:2021kky} for studies of collider bounds.

 \item 14 pNGBs in the electroweak sector in the $\bf{14}$ of $\SO(5)$, which decomposes into 
   $\bf{3}_1+\bf{3}_0+\bf{3}_{-1}+\bf{2}_{1/2}+\bf{2}_{-1/2}+\bf{1}_0$ under $\SU(2)_L\times \U(1)_Y$, with the 4 degrees of freedom in $\bf{2}_{1/2} + \bf{2}_{-1/2}$ identified as the composite Higgs doublet.
  This sector has been studied in \cite{Agugliaro:2018vsu,Banerjee:2022xmu}, to which we refer 
  for further details. The main aspect relevant for the discussion below is that, after the EW symmetry
  breaking, the bi-triplet $\bf{3}_1+\bf{3}_0+\bf{3}_{-1}$ of $\SU(2)_L\times \SU(2)_R$ decomposes into a singlet $\eta_1$, a triplet $\eta_3$ and
  a quintuplet $\eta_5$ of the custodial diagonal $\SU(2)$.

\item 14 pNGBs in the color sector in the $\bf{14}$ of $\Sp(6)$, which decomposes into $\bf{8_0}+\bf{3_{2/3}}+\bf{\bar{3}_{-2/3}}$ under $\SU(3)_c\times \U(1)_Y$. In the following we will refer to $\pi_8$ and $\pi_3$ as octet and triplet pNGBs, respectively, and refer to \cite{Cacciapaglia:2021uqh} for technical details on the embedding of
$\SU(3)_c$ in $\Sp(6)$. 
The $\pi_8$ couples via
 the Wess-Zumino-Witten term to two gluons. Depending on the model details it can also couple to an
 $q\bar{q}$ pair with a strength proportional $m_q/f_\chi$ with $f_\chi$ being the decay constant related
 to this pNGB sector. Its phenomenology has been investigated in ref.~\cite{Cacciapaglia:2020vyf} where also
 mass bounds $m_{\pi_8}\gsim 1.1$~TeV from LHC data have been obtained. We defer possible couplings of $\pi_3$
 after the presentation of the chimera hyper-baryons.

\end{enumerate}

The model contains fermionic resonances (chimera hyper-baryons) in the confined phase corresponding to 
composite operators made of one $\psi$ and two $\chi$ hyper-fermions. They can be classified in terms of 
their transformation properties under the stability group $\SO(5) \times \Sp(6)$, see ref.~\cite{Cacciapaglia:2021uqh} for further details. We focus here on hyper-baryons transforming as \textbf{14} of $\Sp(6)$ which
decomposes under $\SU(3)_c\times \U(1)_X$ as
\begin{align}
\bf{14}& \rightarrow \bf{8_0}+\bf{3_{2/3}}+\bf{\bar{3}_{-2/3}} \,.
\end{align}
Here we have fixed $\U(1)_X$ charge such that the color-triplets can mix with the SM elementary top fields 
to generate partial compositeness for the top quark mass origin. The components of the $\bf 14$ are 
embedded in the anti-symmetric matrix
\begin{equation} 
\Psi_{\bf 14} =  \begin{pmatrix}  -Q_3^{c} & - \frac{1}{2\sqrt{2}} Q_8^{a}\ \lambda^a  \\  \frac{1}{2\sqrt{2}} Q^{a}_8\ (\lambda^a)^T  & -Q_3  \end{pmatrix}\,,
\label{eq:tom_a17}
\end{equation}
where $Q^{(c)}_{3,ij}=\frac{1}{2} \epsilon_{ijk}Q^{(c)}_{3,k}$  gives correct transformations for the $\SU(3)_c$ generator embedding (i.e. the diagonals  transform like $\bf{3}\times \bf{3} \supset \bf{\bar{3}}$ and $\bf{\bar{3}}\times \bf{\bar{3}} \supset \bf{3}$ while the off-diagonals transform like octets). Each component $Q_3$ and $Q_8$ also transform as a fundamental of $\SO(5)$:
\begin{subequations}
\begin{align}
Q_3 &=  (X_{5/3}, X_{2/3},T_L,B_L,iT_R)^T\, , \label{eq:q3compo}\\
Q^{c}_{3} &=  ( B_L^c,-T_L^c, -X^c_{2/3}, X^c_{5/3},-i T_R^c)^T\, , \label{eq:q3ccompo}\\
Q_8 &= (\tilde G^+_u, \tilde G^0_u, \tilde G^0_d, \tilde G^-_d, i \tilde g)^T\,. \label{eq:q8compo}
\end{align}
\end{subequations}
The first four components transform as two doublets of $\SU(2)_L$ (and a bi-doublet of $\SU(2)_L \times \SU(2)_R$), while the fifth is a singlet.

The couplings of the top fields to the hyper-baryons depend on the choice of three-hyper-fermion operator. 
More specifically, the partial compositeness couplings for the left-handed top (in the doublet $q_{L,3}$) 
and right-handed top $t_R^c$ are assumed to originate from a four-fermion interaction such as
\begin{equation} \label{eq:topPC4F}
    \frac{\xi_L}{\Lambda_t^2} \psi \chi \chi q_{L,3}  \quad \text{ and } 
    \quad \frac{\xi_R}{\Lambda_t^2} \psi \chi \chi t_{R}^c 
\end{equation}
with the appropriate components of the hyper-fermions. Here we have chosen to couple them to the operator $\psi \chi \chi$ in the channel $({\bf 5}, {\bf 15})$ of $\SU(5)\times \SU6)$. 
This choice implies the presence of the singlet baryons as well \cite{Cacciapaglia:2021uqh}:
\begin{equation}
Q_1 = (\tilde h^+_u, \tilde h^0_u, \tilde h^0_d, \tilde h^-_d, i \tilde B)^T\,. \label{eq:q1compo}
\end{equation}
We note for completeness that all the hyper-baryon fields introduced so-far are 2-component Weyl spinors.  
Henceforth, the top partners in this model include all the components of the $\bf 14$ and the singlet, 
as all their components couple with the top. 

We see that the low energy spectrum of this model contains unusual top partners, transforming as color octets 
and color singlets,  besides the usual top partners in eqs.~\eqref{eq:q3compo} and \eqref{eq:q3ccompo}. They 
correspond to the following states, written in terms of 4-components spinors:
\begin{subequations}\label{eq:hyperbaryonnames}
\begin{align}
\mbox{Octoni (Dirac): \hspace{10pt}} \tilde G^+ &= \begin{pmatrix} \tilde G^+_u \\ \bar{\tilde {G}}^-_d \end{pmatrix} \,\,,\,\,
\tilde G^0 = \begin{pmatrix} \tilde G^0_u \\ \bar{\tilde {G}}^0_d \end{pmatrix}\,\,, \quad
&\mbox{Gluoni (Majorana): \hspace{10pt}} \tilde g &= \begin{pmatrix} \tilde g \\ \bar{\tilde{g}} \end{pmatrix} \,\,; \\
\mbox{Higgsoni (Dirac): \hspace{10pt}}\tilde h^+ &= \begin{pmatrix} \tilde h^+_u \\ \bar{\tilde{h}}^-_d \end{pmatrix} \,\,,\,\,
\tilde h^0 = \begin{pmatrix} \tilde h^0_u \\ \bar{\tilde{h}}^0_d \end{pmatrix} \,\,, \quad &
\mbox{Boni (Majorana): \hspace{10pt}}\tilde B &= \begin{pmatrix} {\tilde B} \\ \bar{\tilde{B}} \end{pmatrix} \,.
\end{align}
\end{subequations}
The indices $u$ and $d$ indicate that the corresponding $\SU(2)_L$ doublets have isospin $1/2$ and
$-1/2$, respectively.
The choice of the names is motivated by the fact that the states $\tilde h$, $\tilde B$ and $\tilde g$ 
have the same quantum numbers as the higgsino, bino, and gluino in supersymmetric extensions of the SM.

The low energy Lagrangian for the hyper-baryons can be worked out following the Coleman-Callan-Wess-Zumino 
prescription \cite{Coleman:1969sm,Callan:1969sn}. This prescription needs to be adapted along the lines
of \cite{Marzocca:2012zn} to include couplings to SM fermions which are implemented in incomplete
representation of the $\SU(5)\times \SU(6)$. The details as well as the resulting Lagrangian are given in 
ref.~\cite{Cacciapaglia:2021uqh}. It turns out that all the couplings of this Lagrangian allow to assign 
SM baryon number charges B to all the top partners, so that B is still preserved in the presence of 
top partial compositeness. This results in all the color triplets to carry $\text{B}=1/3$, like quarks, 
while color octets and singlets remain neutral. Hence,
\begin{equation}
    \text{B} = 1/3\;\; \mbox{for}\;\; Q_3\,,\, \pi_3\,; \qquad \text{B} = 0\;\; \mbox{for}\;\; Q_8\,,\, Q_1\,,\, \pi_8\,. 
\end{equation}
The lightest state among the components of $Q_8,\ Q_1$ and $\pi_3$ must be stable in absence of either baryon or lepton number violation as a consequence, as there is no matching SM final state.

\section{LHC phenomenology}
\label{sec:LHCpheno}

\begin{figure}
	\centering
	\includegraphics[width=0.5\textwidth]{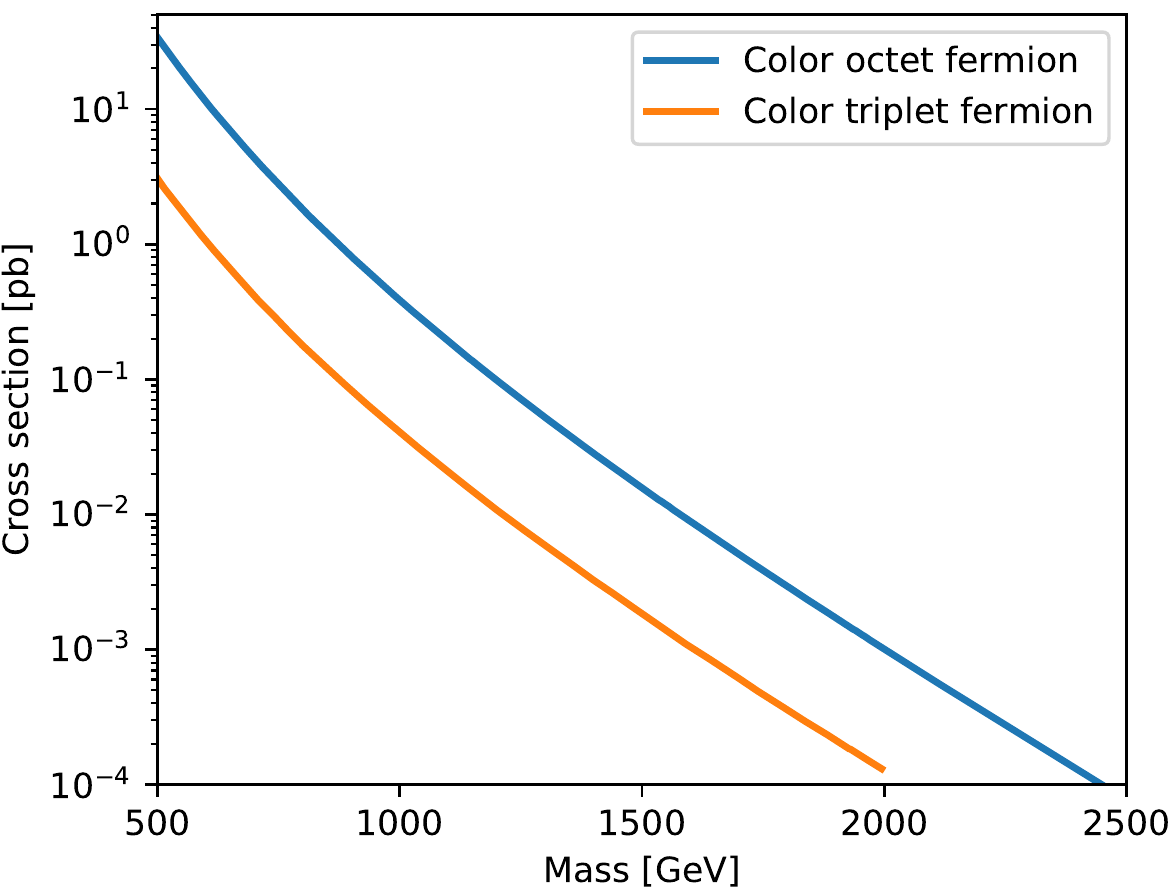}
	\caption{Comparison of the production cross sections of a color triplet top partner (at QCD NLO, from \cite{Fuks:2016ftf}) and a color octet Majorana top partner (at NNLO$_\text{approx}$+NNLL from \cite{Beenakker:2016lwe}).}
	\label{fig:q3q8xs}
\end{figure}

The LHC phenomenology of this model depends crucially on the mass hierarchy among the QCD-colored baryons,
which have the largest production cross sections. They have a common mass
$M_{14}$ as they all belong to the same baryon multiplet -- the $\bf{14}$ of $\Sp(6)$ --  and the 
mass differences are only due to the SM gauge interactions and the top couplings.
The octet top partners, however, enjoy the largest pair production cross section thanks to their QCD quantum numbers, as shown in fig.~\ref{fig:q3q8xs} which is particularly important if the baryons have 
approximately the same mass. 

\begin{figure}[tb]
	\centering
	\includegraphics[width=0.9\textwidth]{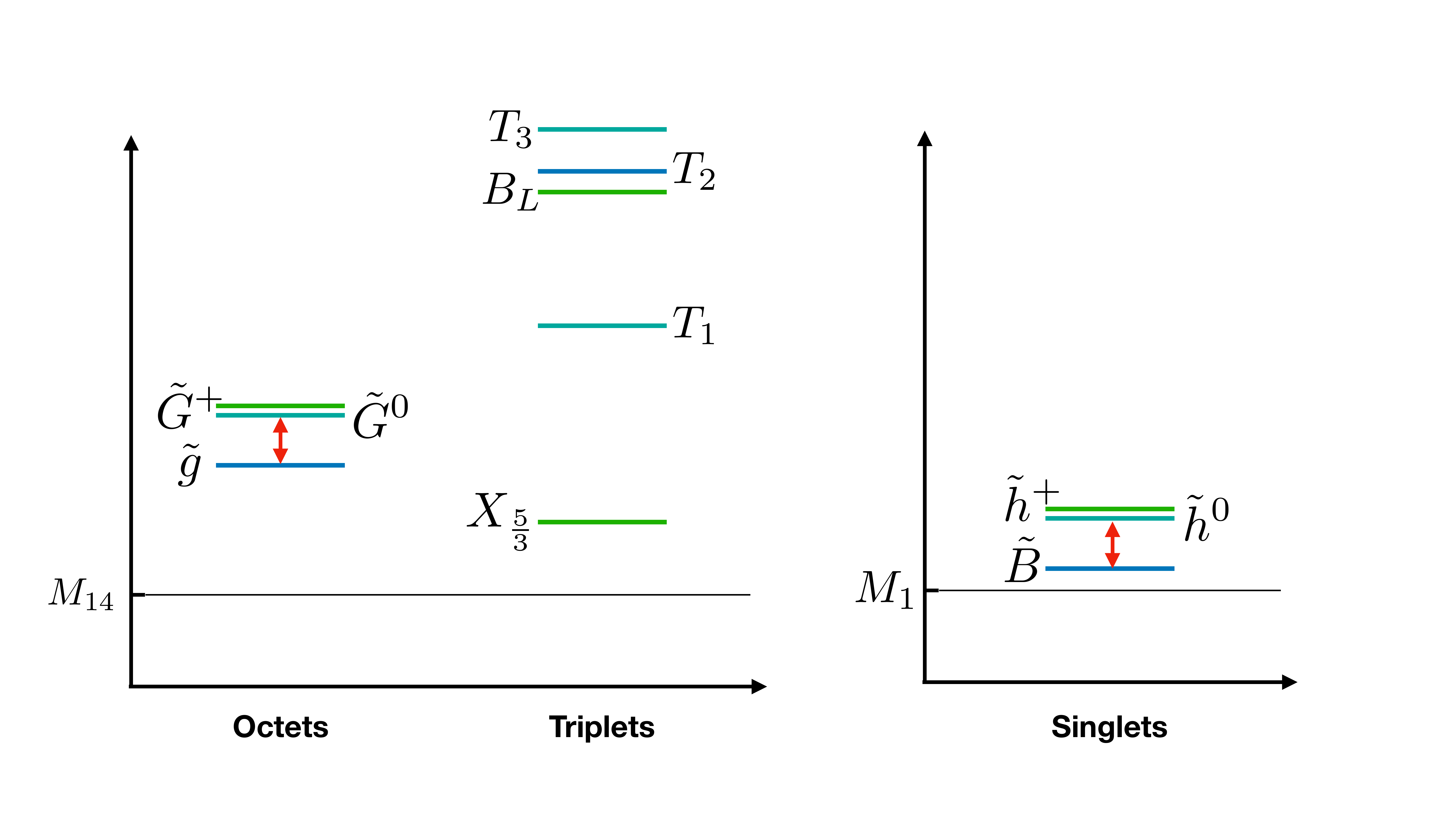}
	\caption{Template spectrum of the chimera baryon states, where $T_{1,2,3}$ ($B_L$) are the mass eigenstates with the same charge as the top (bottom) quark. They receive a positive mass shift due to the mixing with the top quark. The mass splitting highlighted by the red arrow can also receive significant contribution from the different masses in the $\psi$ sector.}
	\label{fig:spectrum}
\end{figure}

A typical spectrum for the colored top partners is illustrated in Fig.~\ref{fig:spectrum} (left).
The triplets that have charges matching the top and bottom quarks, i.e. $2/3$ and $-1/3$, receive 
large positive corrections due to the mixing with the top fields, which makes them heavier than the octets 
and the $X_{5/3}$. The mass difference between the octets and $X_{5/3}$ is due to QCD corrections
and can be been estimated as \cite{Cacciapaglia:2021uqh}
\begin{align}
\frac{2 (m_{\tilde{g}} - m_{X_{5/3}})}{m_{\tilde{g}} + m_{X_{5/3}}} \approx \frac{\alpha_S (\text{TeV})}{\alpha_{em} (\text{GeV})} \left(3   -\frac{4}{3}\right)\ r \sim 1.4\%\,. 
\end{align}
Here we have used $r= 2 \Delta m^{\rm em}/(m_P+m_N) \simeq 6 \cdot 10^{-4}$ with 
$\Delta m^{\rm em} \simeq 0.58$~MeV being the electromagnetic contribution to the proton neutron mass 
difference \cite{Borsanyi:2014jba,Gasser:2020mzy}.
This estimate confirms that the octet top partners are much more abundantly produced at the LHC compared to the usual triplet top partners. 
Note that the mass difference between the octonis and the gluoni is generated by EW corrections, which we estimate to be
\begin{align} \label{eq:weak_splitting}
\frac{2 (m_{\tilde{G}} - m_{\tilde{g}})}{m_{\tilde{G}} + m_{\tilde{g}}} \approx \frac{3}{4} \frac{1}{\sin^2 \theta_W} \ r \sim 0.2\%\,,
\end{align}
where we included the dominant contribution of $\SU(2)_L$. 
We note for completeness that a contribution to this mass difference is also generated by 
$\SU(5)$-breaking mass differences in the $\psi$ sector, which could go in either direction.
Finally, we expect the charged octoni to be slightly heavier than the neutral one due to 
electroweak symmetry breaking effects.

The decay patterns of the gluoni and the octoni depend on the mass hierarchy with the singlet top 
partners and the QCD-colored pNGBs. We expect the pNGBs to be lighter than the colored 
hyper-baryons. However, the singlet top partners receive a mass $M_1 \neq M_{14}$ from the 
confining strong dynamics, and the precise values can only be obtained from lattice studies. 
We will explore two different scenarios for $M_1$.
As discussed above, the lightest state among $\tilde{B}, \tilde{h}^{0,\pm}, \pi_3$ is stable 
unless additional baryon or lepton number violating interactions are added to the model. 
A stable $\tilde{B}/\tilde{h}^0$ provides a potential dark matter (DM) candidate
if either $\tilde{B}$ and/or $\tilde{h}^0$ are lighter than $\pi_3$. In case that $\pi_3$ is 
lightest among these particles, a stable $\pi_3$ is not viable, and new interactions must be present,
which open baryon and/or lepton number violating $\pi_3$ decay channels.  Note that this 
could also occur if the singlet top partners are the lightest. Both cases carry resemblance with
supersymmetric signatures: gluinos decaying into tops plus missing transverse energy in the first 
case and R-parity violating decays in the second one.

\paragraph{Scenarios with a DM candidate:}
In this scenario the singlet top partners,  boni or higgsonis, are lighter than the colored pNGBs, and cannot decay into any SM final state being the lightest hyper-baryons. 
We  note, that the top partners $\tilde h^0$, $\tilde h^+$ and $\tilde B$ and the pNGB 
$\pi_3$ have quantum 
numbers resembling the higgsino-bino and right-handed stop sectors of supersymmetric models.
Their phenomenology depends on the mass hierarchy among them and,
considering only electroweak interactions, one finds
\begin{align}
m_{\tilde h^+} \gtrsim m_{\tilde h^0} >  m_{\tilde B}\,, 
\end{align}
as shown in fig.~\ref{fig:spectrum} (right) with a mass splitting estimated to range around 
$2\cdot 10^{-4}\, M_1$,  see 
eq.~\eqref{eq:weak_splitting}.  A small mixing is also generated by electroweak symmetry breaking. 
However, the mass difference between the boni and the higgsonis also receives a sizeable contribution from the $\SU(5)$-breaking mass differences between the singlet and bi-doublet hyper-fermions 
$\psi$, which could go in either direction. 
The most natural expectation is that the spectrum remains fairly compressed, hence
we would expect $\tilde h^+$ and $\tilde h^0$ to decay into soft leptons and mesons plus $\tilde B$.
Thus, all three particles would effectively contribute to the missing transverse momentum 
at the LHC as
the soft decay products are hardly registered in the detectors. Consequently, such a scenario
might be easily confused with a supersymmetric model at first glance. 

The QCD-colored pNGB $\pi_3$ will decay into a SM-quark and either a boni or higgsoni as it carries baryon number:
\begin{align} \label{eq:pi3decayA}
\pi_3 \to t  \tilde B,\ t \tilde{h}^0,\ b \tilde{h}^+\,.
\end{align}
This resembles the decays of stops in supersymmetric models.
In principle, decays into lighter families, like $c \, \tilde B$ and $u \, \tilde B$, are also 
possible, but in the spirit of composite Higgs models we expect those to be strongly suppressed.
In the following we assume that the decay into $\tilde B\, t$ dominates. In other words,
$\pi_3$ behaves exactly like a right-handed stop in supersymmetry, and LHC bounds from stop  searches can be directly applied \cite{CMS:2019zmd,ATLAS:2020xzu,CMS:2021eha}. For large
mass differences this gives a bound of about 1.3 TeV. 
We note fore completeness, that in case of small mass difference between $\pi_3$ and $\tilde B$
three-body decays via an off-shell top-quark would become important similar to supersymmetric models 
\cite{Porod:1996at,Porod:1998yp,Boehm:1999tr}.

In the scenario considered here, the decays of octet top partners $\tilde G^{0,+}$ and $\tilde g$ 
lead to final states similar to those of gluinos in supersymmetric models. This applies 
in particular to the Majorana gluoni $\tilde g$ that decays via the following channels
\begin{align}
\tilde g &\to \pi_3 \,\bar{t}\,,\,   \pi_3^* t \to t\,\bar{t} \, \tilde B \ / \ t\,\bar{t} \, \tilde{h}^0 + t\,\bar{t} \, \overline{\tilde{h}^0} \ / \ b\,\bar{t} \, \tilde{h}^+ +  t\,\bar{b} \, \tilde{h}^-  \quad \mbox{and/or} \quad %\label{eq:gpi3dec}\\
\tilde g    \to \pi_8 \tilde B \to t\,\bar{t} \, \tilde B\ / \  g\,g \, \tilde B \, ,\label{eq:gpi8dec}
\end{align}
depending on the mass spectrum.
In all cases the final states contain large missing transverse momentum caused by the assumed
stability of $\tilde B$. Consequently, gluino searches at the LHC can be used to constrain these scenarios. 
A similar comment applies in case of the Dirac states $\tilde G^0$ and $\tilde G^+$, which can
decay into the following channels
\begin{subequations}
\begin{align}
\tilde G^0  \to \pi_3 \,\bar{t}  \to t\,\bar{t} \, \tilde B \ / \ t\,\bar{t} \, \tilde{h}^0 \ / \ b\,\bar{t} \, \tilde{h}^+ \quad &\mbox{and/or} \quad 
\tilde G^0  \to \pi_8 \, \tilde h^0   \to t\,\bar{t} \, \tilde h^0 \ /\ g \, g \, \tilde h^0\,;  \label{eq:G0pi8dec}\\
\tilde G^+  \to \pi_3 \,\bar{b}  \to t\,\bar{b} \, \tilde B \ / \ t\,\bar{b} \, \tilde{h}^0 \ / \ b\,\bar{b} \, \tilde{h}^+ \quad & \mbox{and/or} \quad
\tilde G^+  \to \pi_8 \, \tilde h^+   \to t\,\bar{t} \, \tilde h^+ \ / \ g \, g \, \tilde h^+ \,.\label{eq:G+pi8dec}
\end{align} \end{subequations}
We note for completeness, that $\tilde{G}^0$ resembles a Dirac gluino in extended SUSY models 
\cite{Choi:2008pi,Benakli:2014cia}, while the charged octoni $\tilde{G}^+$ is a novel state from 
composite models without a supersymmetric analogue.
Moreover, we note that for a fixed mass the QCD production cross sections fulfil the relation
\begin{align}
\sigma(p\,p\to \tilde G^0 \, \overline{\tilde G^0}) = \sigma(p\,p\to \tilde G^+ \, \tilde G^-)
= 2\, \sigma(p\,p\to \tilde g \, \tilde g)\,.
\end{align}
\vskip 2mm

\paragraph{Scenarios without a DM candidate:} 
If $\pi_3$ is lighter than $\tilde{B}$, lepton or baryon number violating interactions need to be included in order to avoid a stable $\pi_3$. 
The simplest possibilities are 
\begin{align} \label{eq:pi3decayBb}
\pi_3 \to \bar{d}_i \, \bar{d}_j \quad \text{ with } d_i = d,s,b \text{ and } i\ne j 
\end{align}  
or
\begin{align} \label{eq:pi3decayBl}
\pi_3 \to u_i \, \nu_{l_j} \,,\, d_i \, l_j \quad \text{ with } u_i = u,c,t \text{ and } 
l_j = e,\mu,\tau\,. 
\end{align}  
The former violates baryon number whereas the latter violates lepton number.
Note, that only one of the two interaction types can be present as otherwise the  proton could
decay at a rate incompatible with experiment. This scenario corresponds to typical R-parity
violating supersymmetric models for  stop decays. However, we stress that the origin of 
these couplings is very different from the supersymmetric case and no R-parity analogue
exists in composite models.

The QCD-singlet top partners $\tilde h^0$, $\tilde h^+$ and $\tilde B$ can decay according
to 
\begin{align}
\tilde B &\to \pi^*_3 \, t \,,\, \pi_3 \, \bar{t}\,; \quad
\tilde h^0  \to \pi_3  \, \bar{t}\,; \quad \text{ and }\quad 
\tilde h^+  \to \pi_3  \, \bar{b} \,.
\end{align}
Moreover, there could be mixing of $\tilde h^0$ and $\tilde h^-$ with the left-handed leptons,
extending the spirit of partial compositeness to the lepton sector. 
 Additional decay channels into 
electroweak gauge bosons and pNGBs would be present, such as
\begin{subequations}\begin{align}
\tilde h^0 & \to Z \, \nu \,,\, W^+ \, l^- \,,\, h\, \nu \, ,\, \eta^+_{3,5} l^- \\
\tilde h^+ & \to W^+ \, \nu \,,\, Z \, l^+ \,,\, h\, l^+ \, ,\, \eta^+_{3,5} \nu \\
\tilde B & \to h\, \nu \, ,\, \eta^\pm_{3,5} l^\mp  \,,
\end{align}\end{subequations}
to name a few. Note that this possibility is only compatible with the $\pi_3$ decay in 
eq.~\eqref{eq:pi3decayBl}, as it involves lepton number violation, and it also holds 
if $\pi_3$ is heavier than these states. As a consequence, final states from the decays of 
the color-octet baryon will contain additional jets and leptons from the new decays 
of $\pi_3$ and the singlet baryons, and reduced missing transverse momentum.

\subsection{LHC bounds on color octet hyper-baryons}
\label{sec:numerics}

We concentrate here on the case with a stable $\tilde B$ as in this case one can re-interpret
gluino searches in the present context. We have seen above, that the color-octets are among
the lightest color-charged hyper-baryons. Their cross-section is significantly larger 
compared to those of the color triplets, see fig.~\ref{fig:q3q8xs}.
We assume that all  octet top partners $Q_8=(\tilde{g}, \tilde{G}^{+,0})$ have the same mass. 
Furthermore, we assume the  higgsonis $\tilde{h}^{+,0}$ to be nearly mass degenerate with 
the boni $\tilde B$ and that they decay promptly to $\tilde B$ plus soft leptons and mesons. Therefore 
all (color) singlet top partners $Q_1=(\tilde{B},\tilde{h}^{+,0})$ have a common mass scale. 
Moreover, we assume that $\pi_3$ decays dominantly into $t\tilde B$ for simplicity.

We have implemented the relevant parts of the M5 model \textsc{FeynRules} 
\cite{Alloul:2013bka,Christensen:2009jx,Degrande:2011ua} 
to generate a leading-order \textsc{LO UFO} model as detailed in \cite{Cacciapaglia:2021uqh}. 
We have used \textsc{MADGRAPH5\_AMC@NLO} \cite{Alwall:2014hca} with the LO set of 
\textsc{NNPDF 3.0} of parton densities \cite{NNPDF:2014otw,Buckley:2014ana} in conjunction 
with \textsc{PYTHIA 8} \cite{Sjostrand:2014zea} to produce hadron-level events of pair-produced 
octet top partners with various decay modes. We have simulated events at LO and we have rescaled 
production cross sections for the Majorana gluoni to the NNLO$_\text{approx}$+NNLL result for 
gluinos with the corresponding mass of Ref.\cite{Beenakker:2016lwe}. For octonis 
we have rescale to  twice the gluino cross section, correspondingly.

We have passed the generated signal events to \textsc{MADANALYSIS 5} version 1.8.44 
\cite{Conte:2012fm,Conte:2014zja,Dumont:2014tja,Conte:2018vmg} for detector simulation, 
event reconstruction based on \textsc{DELPHES 3} \cite{deFavereau:2013fsa} and the 
\textsc{FASTJET} \cite{Cacciari:2011ma} implementation of the anti-$k_T$ algorithm 
\cite{Cacciari:2008gp}, and the extractions of $CL_s$ exclusions relative to the ATLAS and CMS 
searches at the LHC with $\sqrt{s}=13$~TeV which are publicly available in the 
\textsc{MADANALYSIS 5 PAD}. The most sensitive available searches for the final states under 
consideration are ATLAS and CMS searches for stops and gluinos 
\cite{CMS:2019zmd,ATLAS:2019vcq,ATLAS:2018yhd}.

\begin{figure}[t]
 \begin{center}
		\includegraphics[width=0.25\textwidth]{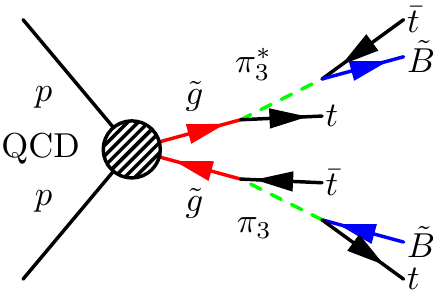} \quad 
		\includegraphics[width=0.25\textwidth]{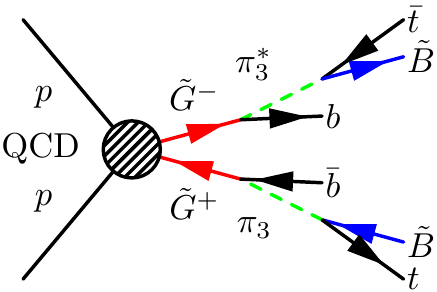} \quad
		\includegraphics[width=0.25\textwidth]{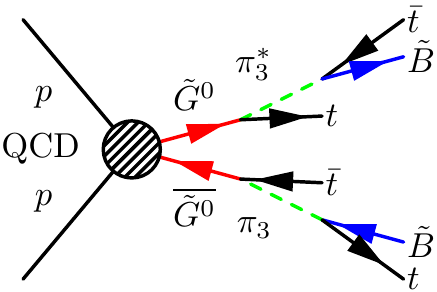} \\[2mm]
		\includegraphics[width=0.25\textwidth]{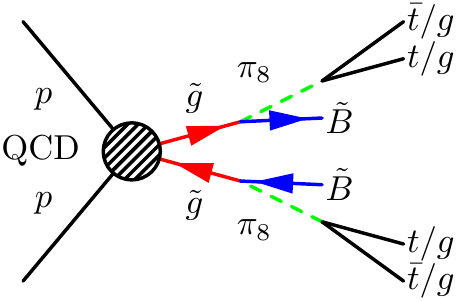}  \quad
		\includegraphics[width=0.25\textwidth]{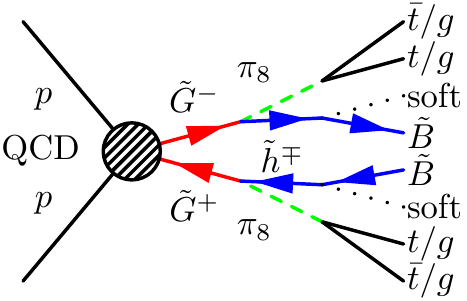}  \quad
		\includegraphics[width=0.25\textwidth]{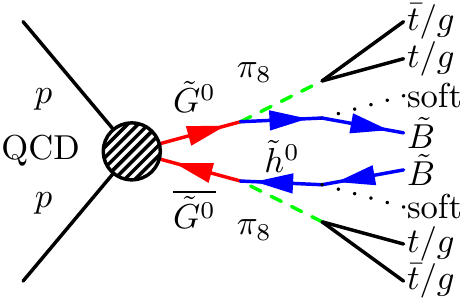}
 \end{center}		
	\caption{Feynman diagrams of QCD pair production of octet top partners $Q_8=(\tilde g,\tilde G^+,\tilde G^0)$ with dominant decays  $Q_8\to\pi_3+t/b$ (upper row) or
	with dominant decays  $Q_8\to Q_1\pi_8$ (lower row).}	
	\label{fig:S1graphs}
\end{figure}

We will focus here on the two limiting cases where the gluoni and octoni either decay
exclusively via a $\pi_3$ or via $\pi_8$. We will also comment on the combination of both
decay possibilities below. The Feynman diagrams for the dominant production cross section combined 
with the corresponding decays are displayed in fig.~\ref{fig:S1graphs}. The signatures are
either $4 t + \ptmiss$, $2 t+ 2j+ \ptmiss$ or $4 j + \ptmiss$ which in case of octoni production
can also be accompanied by soft jets or soft leptons.

The cross section for pair production is solely a function of the octet top partner mass $m_{Q_8}$, but 
the kinematics of the processes depend in addition on the boni mass $m_{Q_1}$ and the mass of 
the involved color pNGB, either $m_{\pi_3}$ or $m_{\pi_8}$, leaving us with 4 relevant mass 
parameters. To present results, we chose three kinematically different setups: 
For setup (i) the mass difference $m_{Q_8}-m_{\pi_3}=200$~GeV
is fixed. 
In this case the $t/b$ of the $Q_8$ decay has little momentum in the $Q_8$ rest-frame. 
For setup (ii) we fix $m_{\pi_3}=1.4$~TeV such that present bounds from stop searches
are satisfied \cite{CMS:2019zmd,ATLAS:2020xzu,CMS:2021eha}.
In setup (iii) we consider decays into $\pi_8$ with $m_{\pi_8}=1.1$~TeV  which is at the 
level of current experimental constraints on $m_{\pi_8}$ \cite{Cacciapaglia:2020vyf}. 
In this setup we consider two limiting cases, namely that the $\pi_8$ decays either
solely into $gg$ or solely into $t\bar{t}$. We scan in all setups over $m_{Q_8}$ and $m_{Q_1}$.

\begin{figure}[th]
\begin{center}
\includegraphics[width=0.35\textwidth]{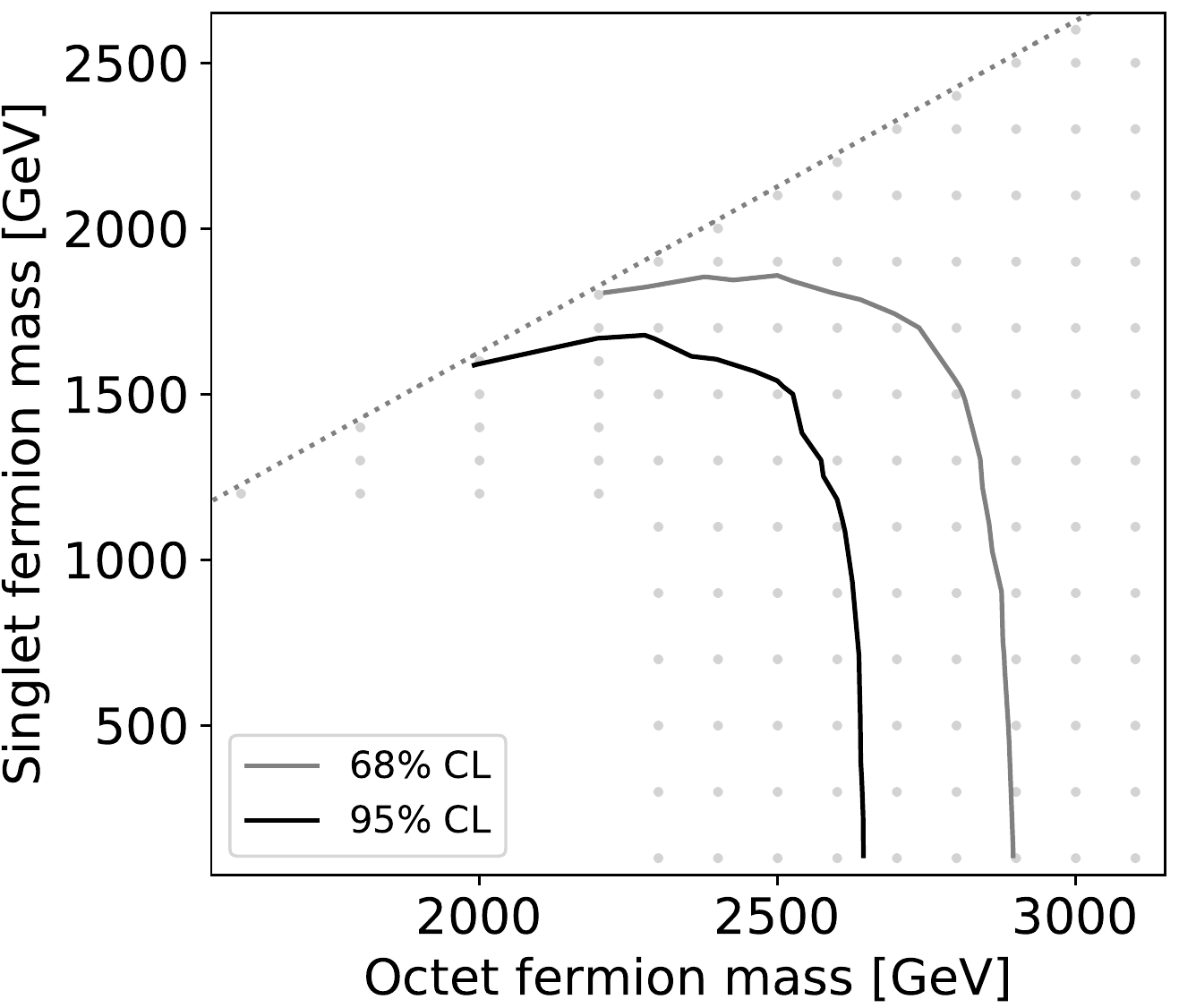} \quad \quad
\includegraphics[width=0.35\textwidth]{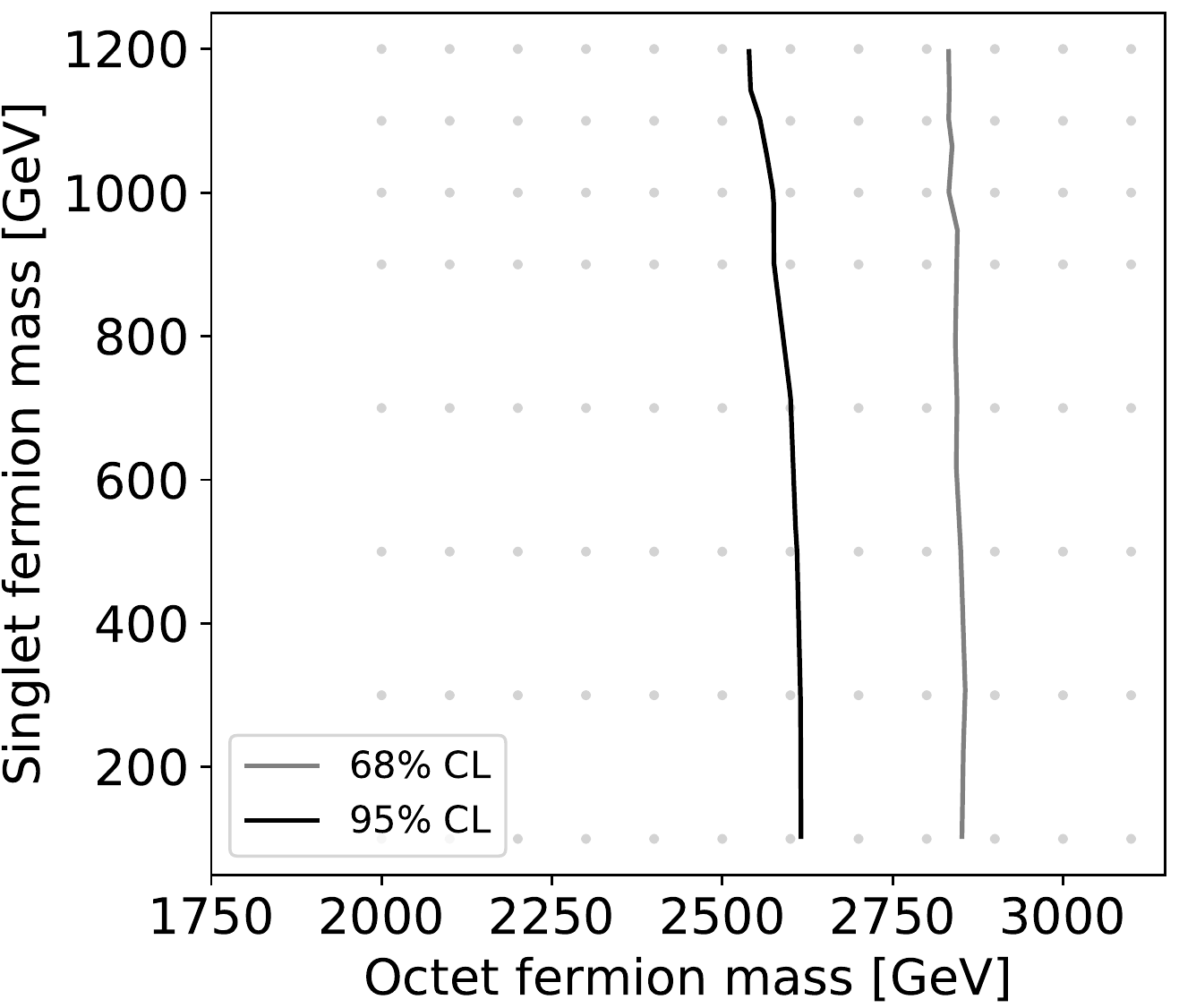} \\
\hspace*{10mm} setup (i)  \hspace{47mm} setup (ii) \\[2mm]
\includegraphics[width=0.35\textwidth]{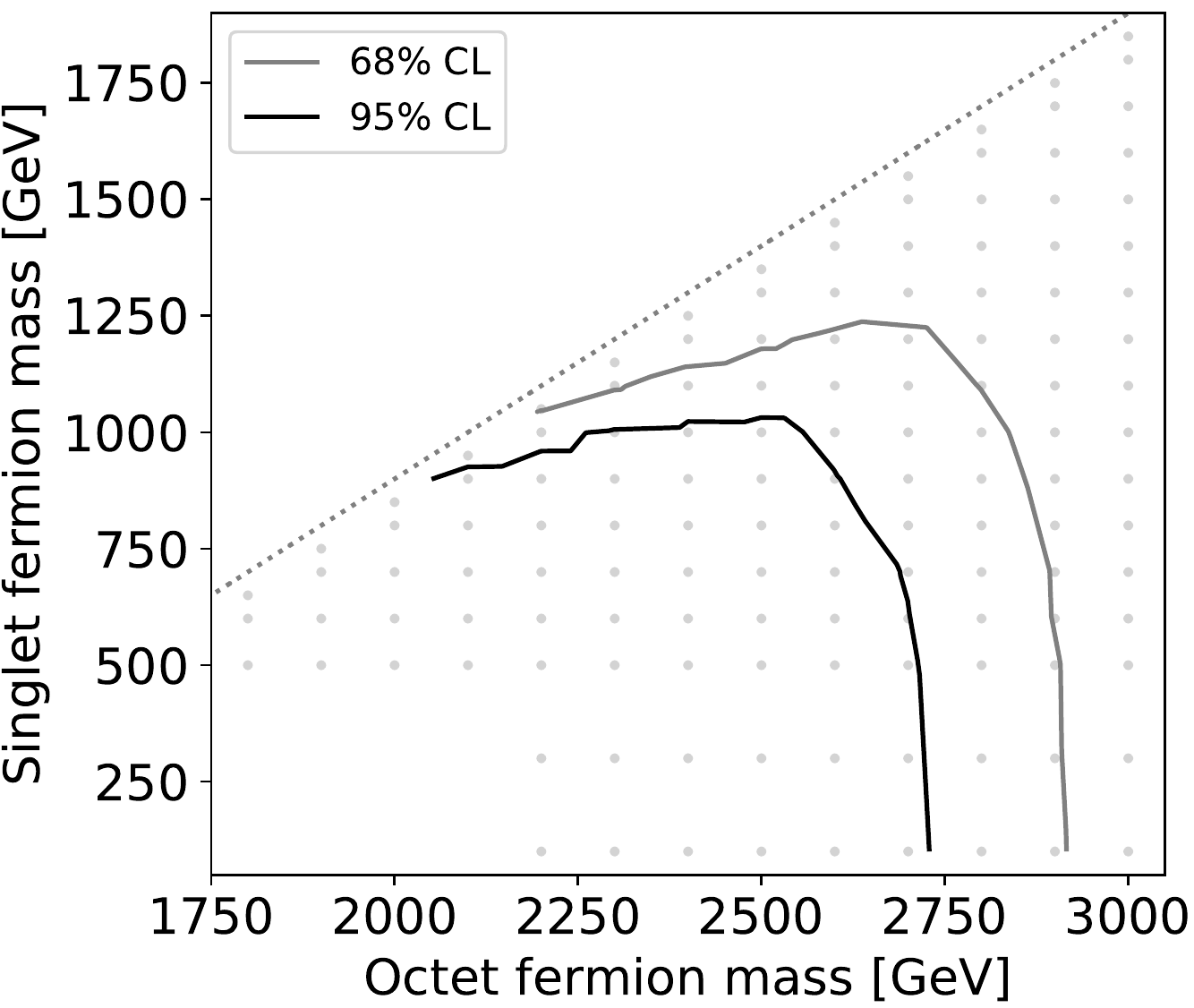} \quad \quad
\includegraphics[width=0.35\textwidth]{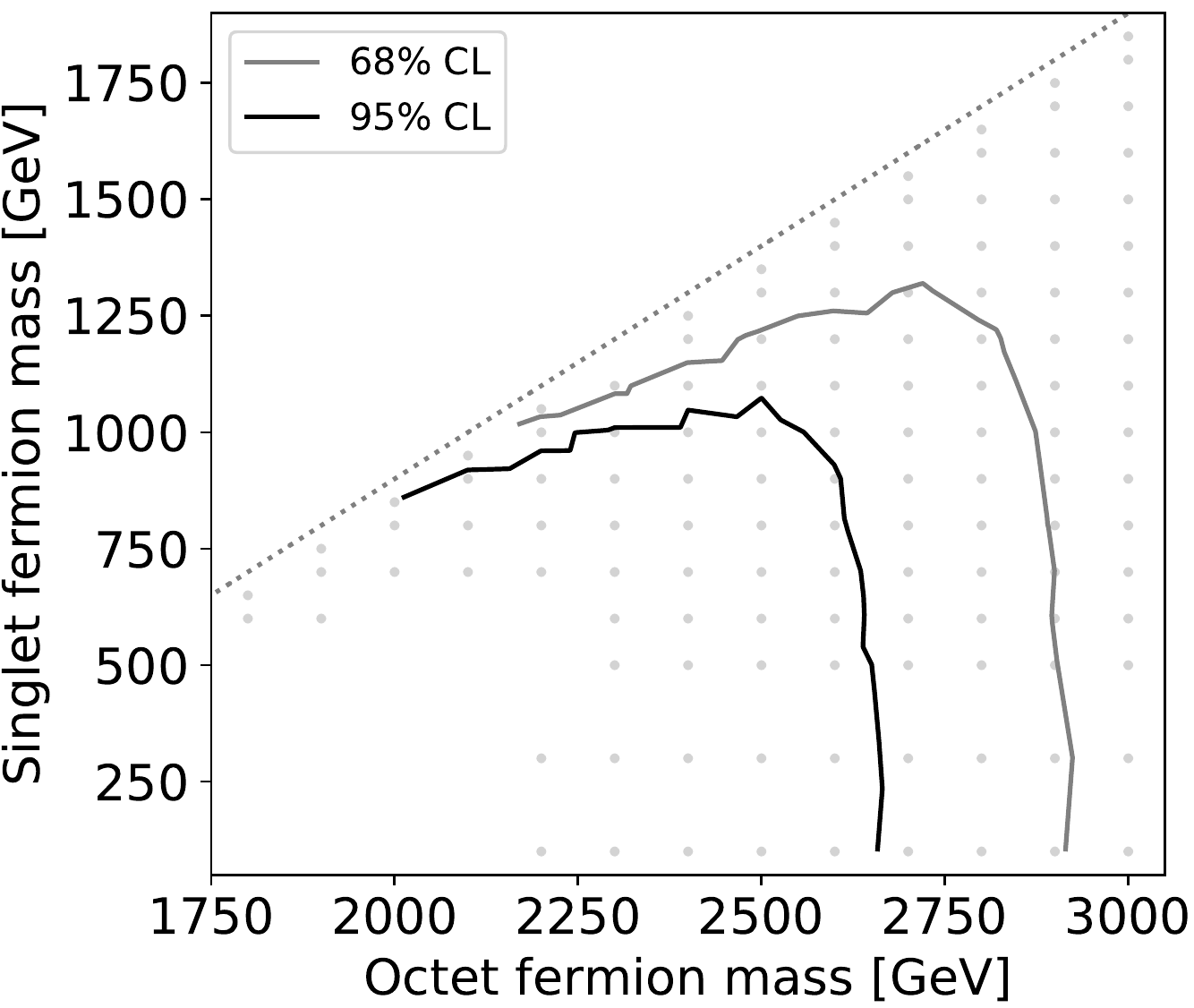}  \\
\hspace*{12mm}setup (iii) with $\pi_8\to gg$ \hspace{22mm} setup (iii) with $\pi_8\to t\bar{t}$
\end{center}
	\caption{Bounds on the fermion masses for QCD pair production of an octet fermion $Q_8$
	in the $m_{Q_8}$-$m_{Q_1}$ plane for 
	different setups: setup (i) $Q_8$ subsequent decay to a SM third generation quark $t/b$ and 
	$\pi_3$ with $m_{Q_8}-m_{\pi_3}=200$~GeV; setup (ii)  $Q_8$ subsequent decay 
	to a SM third generation quark $t/b$ and $\pi_3$ with fixed $m_{\pi_3}=1.4$~TeV;
	setup (iii) $Q_8$ decaying to $\pi_8$ with either $\pi_8\to gg$ (left) or 
	$\pi_8\to t\bar{t}$ (right) and fixed $m_{\pi_8}=1.1$~TeV. 
   Plots are taken from ref.~\cite{Cacciapaglia:2021uqh}.}
	\label{fig:bounds}
\end{figure}

Figure \ref{fig:bounds} shows the resulting bounds for the different cases.
For each scan point, we have generated $10^5$  events which were analysed with LHC searches 
available in \textsc{MADANALYSIS 5}. In most of the parameter space, the leading bounds arise 
from the CMS search focusing on multi jets plus missing transverse momentum  \cite{CMS:2019zmd}.
The black and grey  lines show the obtained 95\% and 68\% CL exclusion boundaries, below which singlet and octet fermion masses are excluded. The upper row of this figure shows the
case where the octet fermions decay into $\pi_3 + t/b$ with $\pi_3 \to t\,\tilde B$. 
We obtain a bound of about 2.65~TeV on the color octet top partner scale $m_{Q_8}$ for light
$\tilde B$.  The lower row shows the case where $Q_8$ decays into $\pi_8 +\ptmiss$ and
$\pi_8$ decays either into $gg$ (left plot) or $t\bar{t}$ (right plot). For light
$\tilde B$ we obtain in 
case of $\pi_8\to gg$ a bound on the octet fermion of about 2.75~TeV and in
case of $\pi_8\to t\bar{t}$ of about 2.7~TeV. This clearly shows that decay products of the 
$\pi_8$ are of lesser importance for this kinematical configuration. However, the plots of the
lower row show clear the impact of the kinematics in case that the octet and singlet hyper-baryons
have close masses. We also note that depend only weakly on the decays of the $Q_8$ and, thus,
one obtains similar bounds in case that both decay channels, $\pi_3$ and $\pi_8$, are of equal
importance, see ref.~\cite{Cacciapaglia:2021uqh} for further details. This finding is non-trivial as the kinematics of the decays are very different.

\section{Conclusions} \label{sec:outlook}

We have explored the collider phenomenology of
composite Higgs models with a fermionic UV completion. As a particular example we have focused on the 
so-called M5 model class. 
A peculiar feature of this model is that the hyper-baryon spectrum contains 
color-octet fermionic states. They are predicted to be among the lightest 
top partners and, thus, play the leading role in the LHC phenomenology of this model. 
We have presented the generic phenomenological features of this model.  We have seen
that, in scenarios where both lepton and baryon number are conserved, the color singlet
top partners can be potentially dark matter candidates. As a consequence the color
octet top partners share several features of gluinos in SUSY models with conserved R-parity. 
This has allowed us to use existing recast tools to obtain mass bounds of 
up to about $2.7$~TeV  on these fermions due to existing LHC analyses. 
The usual color triplet top partners are expected to be in the same mass range 
or even  heavier than the octet baryons, which would be an explanation of 
the null results in the direct LHC searches for these states.

\section*{Acknowledgements}

This work has been supported by the ``DAAD, Frankreich''  PROCOPE 2021-2023, project number 57561441.

\end{document}